\documentclass[12pt]{article}
\usepackage{lineno,hyperref}
\usepackage{amsmath}
\usepackage{amssymb}
\usepackage{graphicx}
\usepackage{esint}

\begin{document}


\title{How to obtain a cosmological constant from small exotic $\mathbb{R}^{4}$}


\author{Torsten Asselmeyer-Maluga \\
German Aerospace Center (DLR), 10178 Berlin, Germany \\
torsten.asselmeyer-maluga@dlr.de \\
and \\
Jerzy Kr{\'o}l\\
Institute of Physics, University of Silesia, 40-007 Katowice, Poland \\
jerzy.krol@us.edu.pl}

\maketitle






\begin{abstract}
In this paper we determine the cosmological constant as a topological
invariant by applying certain techniques from low dimensional differential
topology. We work with a small exotic $R^{4}$ which is embedded into
the standard $\mathbb{R}^{4}$. Any exotic $R^{4}$ is a Riemannian
smooth manifold with necessary non-vanishing curvature tensor. To
determine the invariant part of such curvature we deal with a canonical
construction of $R^{4}$ where it appears as a part of the complex
surface $K3\#\overline{CP(2)}$. Such $R^{4}$'s admit hyperbolic
geometry. This fact simplifies significantly the calculations and
enforces the rigidity of the expressions. In particular, we explain
the smallness of the cosmological constant with a value consisting
of a combination of (natural) topological invariant. Finally, the
cosmological constant appears to be a topologically supported quantity.
\end{abstract}




\section{Introduction}

One of the great mysteries in modern cosmology is the accelerated
expansion of the universe as driven by dark energy. After the measurements
of the Planck satellite (PLANCK) were completed, the model of a cosmological
constant (CC) has been favored among other models explaining the expansion,
like quintessence. In 1917, the cosmological constant $\Lambda$ was
introduced by Einstein (and later discarded) in his field equations
\[
R_{\mu\nu}-\frac{1}{2}g_{\mu\nu}R=\Lambda g_{\mu\nu}
\]
($g_{\mu\nu}$ is a metric tensor, $R_{\mu\nu}$ the Ricci tensor
and $R$ the scalar curvature). By now it seems to be the best explanation
of dark energy. However, the entire mystery of the cosmological constant
lies in its extremely small value (necessarily non-zero, seen as energy
density of the vacuum) which remains constant in an evolving universe
and is a driving force for its accelerating expansion. These features
justify the search for the very reasons explaining their occurrences,
among them the understanding of the small value of the cosmological
constant is particularly challenging. Our strategy in this paper is
to compute the value of a cosmological constant as a topological invariant
in dimension 4.

Such an attempt is far from being trivial or even recognized as possible.
As a motivation to demonstrate the possibility, let us consider the
trace of the Einstein's field equations 
\[
R=-4\Lambda
\]
with a strictly negative but constant scalar curvature for a spacetime.
It follows that the underlying spacetime must be a manifold of constant
negative curvature or admitting an Einstein metric (as solution of
$R_{\mu\nu}=\lambda g_{\mu\nu}$) with negative constant $\lambda=-\Lambda<0$.
The evolution of the cosmos from the Big Bang up to now determines
a spacetime of finite volume. The interior of this finite-volume spacetime
can be seen as compact manifold with negative Ricci curvature. That
is why the corresponding spacetime manifold is diffeomorphic to the
hyperbolic 4-manifold (see the appendix \ref{sec:Appendix-Hyperbolic-Mostow}
about Mostow-Prasad rigidity and hyperbolic manifolds for this uniqueness
result). By Mostow-Prasad rigidity \cite{Mos:68,Prasad:1973}, every
hyperbolic 4-manifold with finite volume is rigid, i.e. geometrical
expressions like volume, scalar curvature etc. are topological invariants.
Then the discussion above indicates that $\Lambda$ might be a topological
invariant. In fact in this paper we show how to calculate the CC as
a topological invariant based on some features of hyperbolic manifolds
of dimension 3 and 4.

It is a rather well-founded and powerful approach in various branches
of physics to look for the explanations of observed phenomena via
underlying topological invariants. There are many examples of such
invariant quantities known from particle physics to solid state physics
as well from the history of physics. Let us mention just two recent
examples, i.e. topological phases in strong electron interactions
and emerging Kondo insulators as heavy fermions \cite{Kondo2017},
or the search for experimental realizations of topological chiral
superconductors with nontrivial Chern numbers (e.g. \cite{Scond2017}).

The distinguished feature of differential topology of manifolds in
dimension 4 is the existence of open 4-manifolds carrying a plenty
of non-diffeomorphic smooth structures. In the computation of the
CC value presented here, the special role is played by the topologically
simplest 4-manifold, i.e. $\mathbb{R}^{4}$, which carries a continuum
of infinitely many different smoothness structures. Each of them except
one, the standard $\mathbb{R}^{4}$, is called \emph{exotic} $R^{4}$.
All exotic $R^{4}$ are Riemannian smooth open 4- manifolds homeomorphic
to $\mathbb{R}^{4}$ but non-diffeomorphic to the standard smooth
$\mathbb{R}^{4}$. The standard smoothness is distinguished by the
requirement that the topological product $\mathbb{R}\times\mathbb{R}^{3}$
is a smooth product. There exists only one (up to diffeomorphisms)
smoothing, the standard $\mathbb{R}^{4}$, where the product above
is smooth. In the following, an exotic $\mathbb{R}^{4}$, presumably
small if not stated differently, will be denoted as $R^{4}$.

But why are we dealing with $R^{4}$? As we mentioned already any
$R^{4}$ (small or big) has necessarily non-vanishing Riemann curvature.
However, the non-zero value of the curvature depends crucially on
the embedding (the curvature is not a diffeomorphism invariant) of
$R^{4}$. That is why our strategy is to look for natural embeddings
of exotic ${R}^{4}$'s in some manifold $M^{n}$ and estimate the
corresponding curvature of this $R^{4}$. This curvature depends on
the embeddings in general. However, we can try to work out an invariant
part of this embedded ${R}^{4}$. If we are lucky enough we will be
able to construct the invariant part of $R^{4}$ (with respect to
some natural embeddings into certain 4-manifold $M^{4}$) with the
topologically protected curvature. We would expect that this curvature
would reflect the realistic value of CC for some (canonical) $M^{4}$.

There are canonical 4-manifolds into which some exotic ${R}^{4}$
are embeddable. Here we will use the defining property of \emph{small}
exotic ${R}^{4}$: every small exotic $R^{4}$ is embeddable in the
the standard $\mathbb{R}^{4}$ (or in $S^{4}$). We analyze these
embeddings in Secs. \ref{sec:How-to-embed} and \ref{sec:Geometric-properties-of}.
There exists a chain of 3-submanifolds of $R^{4}$ $Y_{1}\to\cdots\to Y_{\infty}$
and the corresponding infinite chain of cobordisms 
\[
End(R^{4})=W(Y_{1},Y_{2})\cup_{Y_{2}}W(Y_{2},Y_{3})\cup\cdots
\]
where $W(Y_{k},Y_{k+1})$ denotes the cobordism between $Y_{k}$ and
$Y_{k+1}$ so that $R^{4}=K\cup_{Y_{1}}End(R^{4})$ where $\partial K=Y_{1}$.
The $End(R^{4})$ is the invariant part of the embedding $R^{4}\subset\mathbb{R}^{4}$
mentioned above. In the first part of the paper we will show that
the embedded $R^{4}$ admits a negative curvature, i.e. it is a hyperbolic
4-manifold. This follows from the fact that $Y_{k},k=1,2,...$ are
3-manifolds embedded into hyperbolic 4-cobordism $W(Y_{k},Y_{k+1})$
and the curvature of $Y_{k+1}$, $curv(Y_{k+1})$, is determined by
the curvature of $Y_{k}$: 
\[
curv(Y_{k+1})=curv(Y_{k})\exp(-2\theta)=\frac{1}{L^{2}}\exp(-2\theta)
\]
where $L$ is the invariant length of the hyperbolic structure of
$Y_{k}$ induced from $W(Y_{k},Y_{k+1})$, i.e. $L^{3}=vol(Y_{k+1})$,
and $\theta$ is the topological parameter $\theta=-\frac{3}{2CS(Y_{k+1})}$.
The induction over $k$ leads to the expression for the constant curvature
of the cobordism $W(Y_{1},Y_{\infty})$ as the function of $CS(Y_{\infty})$
\[
curv(W(Y_{1},Y_{\infty}))=\frac{1}{L^{2}}\exp\big(-\frac{3}{CS(Y_{\infty})}\big).
\]
This is precisely what we call the cosmological constant of the embedding
$R^{4}\subset\mathbb{R}^{4}$. It is the topological invariant. However,
in case of the embedding into the standard $\mathbb{R}^{4}$, $Y_{\infty}$
is a (wildly embedded) 3-sphere and thus its Chern-Simons invariant
vanishes. This leads to the vanishing of the cosmological constant
as far as the embedding into $\mathbb{R}^{4}$ is considered. We should
work more globally, namely $R^{4}\subset\mathbb{R}^{4}\subset M^{4}$
and look for the suitable (still canonical) $M^{4}$ into which $R^{4}$
embeds and the corresponding cosmological constant of the cobordisms,
determined by the embedding, assumes realistic value. Now the discussion
is along the line of the argumentation at the beginning of the section
where we considered a hyperbolic geometry on certain Einstein manifold.
At first, one could think that it is not possible at all, that such
miracle can happen, and one can find suitable $M^{4}$ giving the
correct value of CC, and even it can, this $M^{4}$ could not be canonical.
A big surprise of Sec. \ref{real-cc} is the existence of canonical
$M^{4}$, which is the $K3\#\overline{\mathbb{C}P^{2}}$ where $K3$
is the elliptic surface $E(2)$, such that the embedding into it of
certain (also canonical) small exotic ${R}^{4}$, generates the realistic
value of CC as the curvature of the hyperbolic cobordism of the embedding.
Again the curvature is constant which is supported by hyperbolic structure
and thus it can be a topological invariant. In this well recognized
case the boundary of the Akbulut cork of $K3\#\overline{\mathbb{C}P^{2}}$
lies in the compact submanifold $K$ generating $R^{4}$. The boundary
is a certain homology 3-sphere (Brieskorn sphere $\Sigma(2,5,7)$)
which is neither topologically nor smoothly $S^{3}$, contrary to
the previously considered case of the embedding into the standard
$\mathbb{R}^{4}$, and the CS invariant of $\Sigma(2,5,7)$ does not
vanish. Exotic $R^{4}$ as embedded in $K3\#\overline{\mathbb{C}P^{2}}$
lies between this Brieskorn sphere and the sum of two Poincare spheres
$P\#P$. Thus, starting from the 3-sphere (wildly embedded) in $K$
and fixing the size of $S^{3}$ to be of the Planck length, the subsequent
two topology changes take place which allow for the embedding $R^{4}\to K3\#\overline{\mathbb{C}P^{2}}$.
Namely 
\[
S^{3}\to\Sigma(2,5,7)\to P\#P.
\]
Now the ratio of the curvature of the (wildly embedded) $S^{3}$ and
the curvature of $P\#P$ is a topological invariant. Still there is
a freedom to include quantum corrections to this expression. The corrections
are also represented by topological invariants (Pontryagin and Euler
classes of the Akbulut cork). The numerical calculations of the resulting
invariant show a good agreement with the Planck result for the dark
energy density. All details are presented in Sec. \ref{real-cc}.

Some of the material seems to be very similar to our previous work
\cite{AsselmeyerKrol2014}. Therefore we will comment about the differences
between \cite{AsselmeyerKrol2014} and this work. Main idea of \cite{AsselmeyerKrol2014}
is a new description of the inflation process by using exotic smoothness.
Then, inflation as a process is generated by a change in the spatial
topology. In particular we studied a model with two inflationary phases
which will produce a tiny cosmological constant (CC). But the approach
in the paper misses many important points: it was never shown why
CC is a constant, the model uses a very special Casson handle (so
that the attachment is the sum of two Poincare spheres) and it assumed
the embedding of the Akbulut cork in the small exotic $\mathbb{R}^{4}$.
With the results of this paper, these arbitrary assumptions will be
no longer needed. CC is really a constant and we will present the
reason for the constancy (the Mostow-Prasad rigidity of the spacetime).
The model is natural, i.e. there are topological changes starting
with the 3-sphere to Brieskorn sphere $\Sigma(2,5,7)$ and finally
the change to the sum of two Poincare spheres. In contrast to \cite{AsselmeyerKrol2014},
there is no freedom for other topology changes in this paper. Part
of the previous work is the calculation of expansion factor which
was identified with CC. The previous calculation depends strongly
on the embedding. In this paper we will use a general approach via
hyperbolic geometry which will produce a generic result identical
to the previous work. Therefore, some results of this paper are similar
to the previous work but obtained with different methods for a more
general case. We will comment on it in the last two sections.

Secondly, we have to comment about the relation between causality
and topology change in our model. As shown by Andersen and DeWitt
\cite{AndersonDeWitt:1986}{} the singularities of the spatial topology
change imply infinite particle and energy production under reasonable
laws of quantum field propagation. Here, the concept of causal continuity
is central. Causal continuity of a spacetime means, roughly, that
the volume of the causal past and future of any point in the spacetime
increases or decreases continuously as the point moves continuously
around the spacetime. In a series of papers Sorkin, Dowker et.al.
\cite{DowkerGarcia:1998,BordeDowkerGarciaSorkinSurya:1999,DowkerGarciaSurya:2000}
analyzed possible topology changes. In particular they showed that
causal discontinuity occurs if and only if the Morse index is $1$
or $n-1$, i.e. the 4D spacetime has to contain 1- and 3-handles in
its description. By a result of Laudenbach and Poenaru \cite{LaudenbachPoenaru:1972},
the number of 3-handles is determined by the smoothness structure.
For $R^{4}$, no 3-handles are needed \cite{Tay:97}. Furthermore,
by a method of Akbulut \cite{Akbulut:1977} any 1-handle can be described
by removing a 2-handle. Therefore, for our spacetime $R^{4}$ there
is no causal discontinuity. Secondly, the spacetime is topologically
trivial (homeomorphic to $\mathbb{R}^{4}$ with end $S^{3}\times\mathbb{R}$).
Any closed time-like curve will be canceled by a continuous transformation.

Finally, let us comment briefly on the physical meaning of the embedding
of exotic $R^{4}$ into some `big' 4-manifold like $E(2)\#\overline{\mathbb{C}P^{2}}$.
The observed local part of the universe allows for embeddings like
$R^{4}\subset\mathbb{R}^{4}$. However, a more global picture can
exhibit the embeddings like $R^{4}\subset E(2)\#\overline{\mathbb{C}P^{2}}$
as having observed consequences and explaining the CC value. The factor
$\overline{\mathbb{C}P^{2}}$ in 4-dimensional topology of manifolds
is distinguished by itself (blowing up process). However, $\overline{\mathbb{C}P^{2}}$
can be given independent meaning relating quantum field theory contributions
of $R^{4}$ into CC. In the case of the embedding $R^{4}\subset\mathbb{R}^{4}$
the $\overline{\mathbb{C}P^{2}}$ factor disappears and the QFT effects
of $R^{4}$ detected in the ambient standard $\mathbb{R}^{4}$ disappear
either. This is the case of the vanishing of the curvature contributions
to CC, derived in Sec. \ref{sec:Geometric-properties-of} from the
embedding $R^{4}\to\mathbb{R}^{4}$. The QFT contributions are thus
canceled topologically. One recovers these 'QFT contributions' just
by adding the $\overline{\mathbb{C}P^{2}}$ factor. Such an enlarging
the target manifold of the embedding of $R^{4}$ reproduces the correct
value of the vacuum energy density, which is precisely the result
of Sec. \ref{real-cc}. We close the main body of the paper with the
brief discussion of the obtained results. There are three appendixes
attached explaining hyperbolic manifolds together with Mostow-Prasad
rigidity, the concept of a wild embedding and two models (Poincare
disk and half-space model) of hyperbolic geometry used in the paper.

We strongly acknowledge the critical remarks of the anonymous referee.
The response to these remarks increases significantly the readability
of the paper.

\section{Small exotic $\mathbb{R}^{4}$\label{sec:Small-exotic}}

In 4-manifold topology \cite{Fre:82}, a homotopy-equivalence between
two compact, closed, simply-connected 4-manifolds implies a homeomorphism
between them (so-called h cobordism). But Donaldson \cite{Don:87}
provided the first smooth counterexample, i.e. both manifolds, being
h-cobordant, are generally non-diffeomorphic each to the other. The
failure can be localized in some contractible submanifold (Akbulut
cork) so that an open neighborhood of this submanifold is a small
exotic ${R}^{4}$. The whole procedure implies that this exotic ${R}^{4}$
can be embedded in the 4-sphere $S^{4}$. Below we will discuss more
details of the construction.

To be more precise, consider a pair $(X_{+},X_{-})$ of homeomorphic,
but non-diffeomorphic, smooth, closed, simply-connected 4-manifolds.
The transformation from $X_{-}$ to $X_{+}$ can be described by the
following construction. \\
 \emph{Let $W$ be a smooth h-cobordism between closed, simply connected
4-manifolds $X_{-}$ and $X_{+}$. Then there is an open subset $U\subset W$
homeomorphic to $[0,1]\times{{\mathbb{R}}^{4}}$ with a compact subset
$K\subset U$ such that the pair $(W\setminus K,U\setminus K)$ is
diffeomorphic to a product $[0,1]\times(X_{-}\setminus K,U\cap X_{-}\setminus K)$.
The subsets $R_{\pm}=U\cap X_{\pm}$ (homeomorphic to ${{\mathbb{R}}^{4}}$)
are diffeomorphic to open subsets of ${{\mathbb{R}}^{4}}$. Since
$X_{-}$ and $X_{+}$ are non-diffeomorphic, there is no smooth 4-ball
in $R_{\pm}$ containing the compact set $Y_{\pm}=K\cap R_{\pm}$,
so both $R_{\pm}$ are exotic ${R}^{4}$'s.} \\
 Thus, first remove a certain contractible, smooth, compact 4-manifold
$Y_{-}\subset X_{-}$ (called an Akbulut cork) from $X_{-}$, and
then re-glue it by an involution of $\partial Y_{-}$, i.e. a diffeomorphism
$\tau:\partial Y_{-}\to\partial Y_{-}$ with $\tau\circ\tau=Id$ and
$\tau(p)\not=\pm p$ for all $p\in\partial Y_{-}$. This argument
was modified above so that it works for a contractible {\em open}
subset $R_{-}\subset X_{-}$ with similar properties, such that $R_{-}$
will be an exotic ${R}^{4}$ if $X_{+}$ is not diffeomorphic to $X_{-}$.
Furthermore $R_{-}$ lies in a compact set, i.e. a 4-sphere and $R_{-}$
is a small exotic ${R}^{4}$. Freedman and DeMichelis \cite{DeMichFreedman1992}
constructed a continuous family of small exotic ${R}^{4}$'s.

\section{How to embed small exotic ${R}^{4}$ into the standard $\mathbb{R}^{4}$\label{sec:How-to-embed}}

In this section we will construct the embedding of the exotic $R^{4}$
into the standard $\mathbb{R}^{4}$ as well the sequence of non-trivial
3-manifolds $Y_{1}\to\cdots\to Y_{\infty}$ characterizing the exotic
$R^{4}$. This section is a little bit technical and all readers who
accept these facts can switch to the next section.

One of the characterizing properties of an exotic ${R}^{4}$, which
is present in all known examples, is the existence of a compact subset
$K\subset R^{4}$ which cannot be surrounded by any smoothly embedded
3-sphere (and homology 3-sphere bounding a contractible, smooth 4-manifold),
see sec. 9.4 in \cite{GomSti:1999} or \cite{Ganzel2006}. The topology
of this subset $K$ depends strongly on the $R^{4}$. In the example
below, $K$ is constructed from the Akbulut cork of the compact 4-manifold
$E(2)\#\overline{\mathbb{C}P^{2}}.$ Let $\mathbf{R}^{4}$ be the
standard $\mathbb{R}^{4}$ (i.e. $\mathbf{R}^{4}=\mathbb{R}^{3}\times\mathbb{R}$
smoothly) and let $R^{4}$ be a small exotic ${R}^{4}$ with compact
subset $K\subset R^{4}$ which cannot be surrounded by a smoothly
embedded 3-sphere. Then every completion $\overline{N(K)}$ of an
open neighborhood $N(K)\subset R^{4}$ of $K$ is not bounded by a
smooth embedded 3-sphere $S^{3}\not=\partial\overline{N(K)}$. But
$R^{4}$ being small, allows for a smooth embedding $E:R^{4}\to\mathbf{R}^{4}$
in the standard $\mathbf{R}^{4}$. Then the completion of the image
$\overline{E(R^{4})}$ has the boundary $S^{3}=\partial\overline{E(R^{4})}$
as subset of $\mathbf{R}^{4}$. So, we have the strange situation
that an open subset of the standard $\mathbf{R}^{4}$ represents a
small exotic $R^{4}$.

\begin{figure}
\centering \resizebox{0.25\textwidth}{!}{ \includegraphics{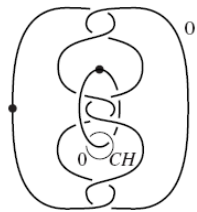}
} 

\caption{link picture for the compact subset $K$ \label{fig:link-picture-for-K}}
\end{figure}

Now we will describe the construction of this exotic $R^{4}$. Historically
it emerged as a counterexample of the smooth h-cobordism theorem \cite{Don:87,BizGom:96}.
The compact subset $K$ as above is given by a non-canceling 1-/2-handle
pair. Then, the attachment of a Casson handle $CH$ cancels this pair
only topologically. A Casson handle is is a 4-dimensional topological
2-handle constructed by an infinite procedure. In this process one
uses disks with self-intersections (so-called kinky handles) and arrange
them along a tree $T_{CH}$: every vertex of the tree is the kinky
handle and the number of branches in the tree are the number of self-intersections.
Freedman \cite{Fre:82} was able to show that every Casson handle
is topologically the standard open 2-handle $D^{2}\times\mathbb{R}^{2}$.
As the result to attach the Casson handle $CH$ to the subset $K$,
one obtains the topological 4-disk $D^{4}$ with interior $\mathbf{R}^{4}$
o the 1-/2-handle pair was canceled topologically. The 1/2-handle
pair cannot cancel smoothly and a small exotic $R^{4}$ must emerge
after gluing the $CH$. It is represented schematically as $R^{4}=K\cup CH$.
Recall that $R^{4}$ is a small exotic ${R}^{4}$, i.e. $R^{4}$ is
embedded into the standard $\mathbf{R}^{4}$, and the completion $\bar{R}^{4}$
of $R^{4}\subset\mathbf{R}^{4}$ has a boundary given by certain 3-manifold
$Y_{r}$. One can construct $Y_{r}$ directly as the limit $n\to\infty$
of the sequence $\left\{ Y_{n}\right\} $ of some 3-manifolds $Y_{n},n=1,2,...$.
To construct this sequence \cite{Ganzel2006}, one represents, by
the use of Kirby calculus of handles, the compact subset $K$ by 1-
and 2-handles pictured by a link say $L_{K}$ where the 1-handles
are represented by a dot (so that surgery along this link gives $K$)
\cite{GomSti:1999}. Then one attaches a Casson handle to this link
\cite{BizGom:96}. As an example see Figure \ref{fig:link-picture-for-K}.

\begin{figure}
\begin{centering}
\resizebox{0.7\textwidth}{!}{ \includegraphics{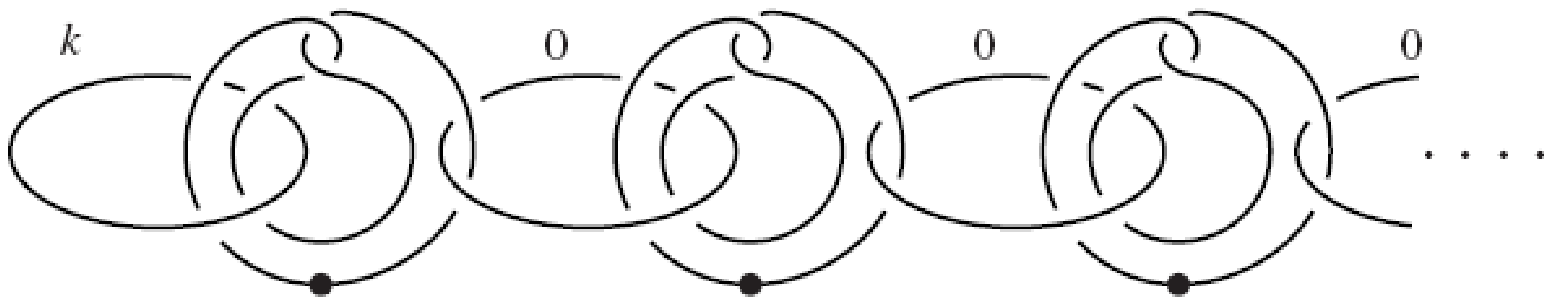}\qquad{}\qquad{}\qquad{}\includegraphics{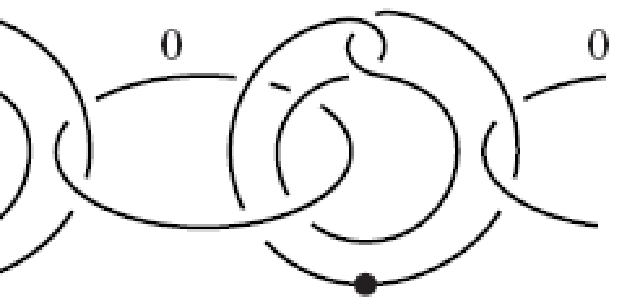}
} 
\par\end{centering}

\caption{building block of every Casson handle (right) and the simplest Casson
handle (left)\label{fig:building-block-simplest-CH}}
\end{figure}

The Casson handle is given by a sequence of Whitehead links (where
the unknotted component has a dot) which are linked according to the
tree (see the right figure of Figure \ref{fig:building-block-simplest-CH}
for the building block and the left figure for the simplest Casson
handle given by the unbranched tree). 

\begin{figure}
\centering \resizebox{0.25\textwidth}{!}{ \includegraphics{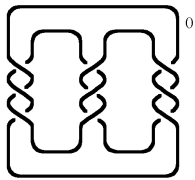}
}

\caption{pretzel knot $(-3,3,-3)$ or the knot $9_{46}$ in Rolfson notation
producing the 3-manifold $Y_{1}$ by $0-$framed Dehn surgery\label{fig:pretzel-knot}}
\end{figure}

For the construction of the 3-manifold $Y_{n}$ which surrounds the
compact $K$, one considers $n-$stages of the Casson handle and transforms
the diagram to a real link, i.e. the dotted components are changed
to usual components with framing $0$. By a handle manipulations one
obtains a knot so that the $n$th (untwisted) Whitehead double of
this knot represents the desired 3-manifold (by surgery along this
knot). Then our example in Figure \ref{fig:link-picture-for-K} with
the $n$-th stages Casson handle, will result in the $n$th untwisted
Whitehead double of the pretzel knot $(-3,3,-3)$ as in Figure \ref{fig:pretzel-knot}
(see \cite{Ganzel2006} for the details of handle manipulations).

Then the entire sequence of 3-manifolds 
\[
Y_{1}\to Y_{2}\to\cdots\to Y_{\infty}=Y_{r}
\]
characterizes the exotic smoothness structure of $R^{4}$. Every $Y_{n}$
is embedded in $R^{4}$ and into $\mathbf{R^{4}}$. An embedding is
a map $i:Y_{n}\hookrightarrow\mathbf{R^{4}}$ so that $i(Y_{n})$
is diffeomorphic to $Y_{n}$. Usually, the image $i(Y_{n})$ represents
a manifold which is given by a finite number of polyhedra (seen as
triangulation of $Y_{n}$). Such an embedding is tame. In contrast,
the limit of this sequence $n\to\infty$ gives an embedded 3-manifold
$Y_{r}$ which must be covered by an infinite number of polyhedra.
Then, $Y_{r}$ is called a wild embedded 3-manifold (see the appendix
\ref{sec:Appendix-wild-emb} and the book \cite{DavermanVenema:2009}).
Then $0-$framed surgery along the pretzel knot for $n=1$ produces
$Y_{1}$ whereas the $n$th untwisted Whitehead double will give $Y_{n}$.
For large $n$, the structure of the Casson handle is coded in the
topology of $Y_{n}$ and in the limit $n\to\infty$ we obtain $Y_{r}$
(which is now a wildly embedded $Y_{r}\subset\mathbf{R}^{4}$ in the
standard $\mathbf{R}^{4}$). But what do we know about the structure
of $Y_{n}$ and $Y_{r}$ in general? The compact subset $K$ is a
4-manifold constructed by a pair of one 1-handle and one 2-handle
which cancel only topologically. The boundary of $K$ is a compact
3-manifold having the first Betti number $b_{1}=1$. This feature
is preserved by taking the limit $n\to\infty$ and characterizes $Y_{r}$
as well. By the work of Freedman \cite{Fre:82}, every Casson handle
is topologically $D^{2}\times\mathbb{R}^{2}$ (relative to the attaching
region) and therefore $Y_{r}$ must be the boundary of $D^{4}$ (the
Casson handle trivializes $K$ to be $D^{4}$), i.e.\emph{ $Y_{r}$
is a wild embedded 3-sphere $S^{3}$}.

\section{Geometric properties of the embedding\label{sec:Geometric-properties-of}}

\begin{figure}
\centering \resizebox{0.6\textwidth}{!}{ \includegraphics{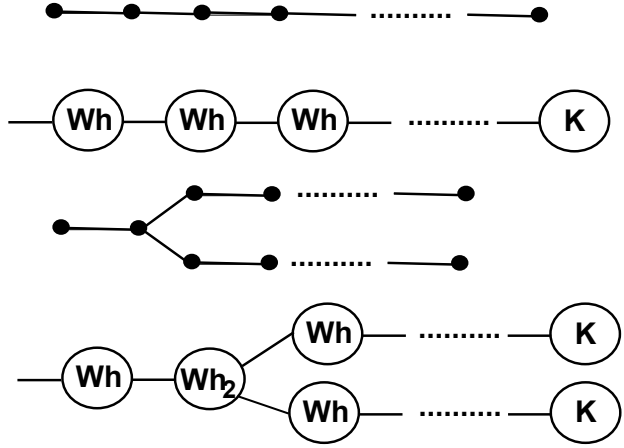}
} \caption{Schematic picture for the splitting of the knot complement $C(\mathcal{K}_{+})$
(above) and in the more general case $C(\mathcal{K_{T}})$ (below)\label{fig:Splitting-of-knot-complement} }
\end{figure}

As we have just seen, the main restriction of the embedding $E:R^{4}\to\mathbf{R}^{4}$
is the sequence of 3-manifolds $Y_{1}\to\cdots\to Y_{\infty}$. $Y_{1}$
was described as the boundary of the compact subset $K$ whereas $Y_{n}$
is given by $0-$framed surgeries along $n$th untwisted Whitehead
double of the pretzel knot $9_{46}$. The entire embedding $E:R^{4}\to\mathbf{R}^{4}$
and the sequence $\{Y_{n}\}$ were determined from the failure of
the h-cobordism and the explicit example of the non-smooth-cobordant
pair of 4-manifolds. Thus we have a sequence of inclusions 
\[
\ldots\subset Y_{n-1}\subset Y_{n}\subset Y_{n+1}\subset\ldots\subset Y_{\infty}
\]
with the 3-manifold $Y_{\infty}$ as limit. Let $\mathcal{K}_{+}$
be the corresponding (wild) knot, i.e. the $\infty$th untwisted Whitehead
double of the pretzel knot $(-3,3,-3)$ ($9_{46}$ knot in Rolfson
notation). The surgery description of $Y_{\infty}$ induces the decomposition
\begin{equation}
Y_{\infty}=C(\mathcal{K}_{+})\cup\left(D^{2}\times S^{1}\right)\qquad C(\mathcal{K}_{+})=S^{3}\setminus\left(\mathcal{K}_{+}\times D^{2}\right)\label{eq:surgery-description-of-Y}
\end{equation}
where $C(\mathcal{K}_{+})$ is the knot complement of $\mathcal{K}_{+}$.
In \cite{Budney2006}, the splitting of the knot complement was described.
Let $K_{9_{46}}$ be the pretzel knot $(-3,3,-3)$ and let $L_{Wh}$
be the Whitehead link (with two components). Then the complement $C(K_{9_{46}})$
has one torus boundary whereas the complement $C(L_{Wh})$ has two
torus boundaries. Now according to \cite{Budney2006}, one obtains
the splitting 
\[
C(\mathcal{K}_{+})=C(L_{Wh})\cup_{T^{2}}\cdots\cup_{T^{2}}C(L_{Wh})\cup_{T^{2}}C(K_{9_{46}}).
\]
This splitting allows us to characterize every right-hand-side factor
(hence left-hand-side either) as a hyperbolic manifold (see Figure
\ref{fig:Splitting-of-knot-complement}). 

At first the knot $K_{9_{46}}$ is a hyperbolic knot, i.e. the interior
of the 3-manifold $C(K_{9_{46}})$ admits a hyperbolic metric. In
general, every link $L$ is hyperbolic if the interior of its complement
$C(L)$ is a hyperbolic 3-manifold (with negative sectional curvature),
see \cite{hyp-knot-theory:2017} for a recent survey. There is a
deep result of Menasco \cite{Menasco:1984}, that every link having
a knot diagram with alternating crossings along each components is
a hyperbolic link if it is non-split (every component is linked) and
prime (it cannot be decomposed into sums) unless it is a torus link.
This theorem can be applied to the knot $9_{46}$ (see Fig. \ref{fig:pretzel-knot}),
i.e. it is an alternating knot and therefore this knot is a hyperbolic
knot. The same argumentation can be used to show that the Whitehead
link is an alternating link as well the $n$th Whitehead link. Using
the result \cite{Budney2006}, every Whitehead double is also a hyperbolic
link because of the splitting of the knot/link complements. The complement
of a Whitehead double of $9_{46}$ is the sum of $C(K_{9_{46}})$
and $C(L_{Wh})$ both admitting a hyperbolic structure. Finally, the
iteration of this case leads to the general case that $C(\mathcal{K}_{+})$
admits also hyperbolic structure, i.e. it is a homogenous space of
constant negative curvature. Therefore we obtained the first condition:
the sequence of 3-manifolds is geometrically a sequence of hyperbolic
3-manifolds.

As a second condition we will remark that the sequence of 3-manifolds
is infinite but the embedding space can be a compact space ($S^{4}$)
or a compact subset of a non-compact space ($\mathbf{R}^{4}$). By
the first condition, we have an infinite sequence of hyperbolic 3-manifolds
with increasing size where the maximal size of $Y_{\infty}$ is bounded
from above. Finally, $Y_{\infty}$ has the topology of a 3-sphere
(using Freedman's famous result). Therefore we will look for a compact
subset with boundary a topological 3-sphere, i.e. a 4-disk $D^{4}$.
The embedding of the infinite sequence into a compact subset enforces
us to choose a 4-ball with metric 
\begin{equation}
ds_{4D}^{2}=\frac{dr^{2}+r^{2}d\Omega^{2}}{(1-r^{2})^{2}}\label{eq:metric-hyp-4-ball}
\end{equation}
with the angle coordinates $\Omega$ (a tupel of 3 angles) and the
radius $r$, i.e. the Poincare hyperbolic 4-ball. At first we will
remark that for a fixed $r=const.$ the corresponding submanifold
in the 4-disk $D^{4}$ is a 3-sphere. The embedding of the sequence
$Y_{1}\to Y_{2}\to\cdots\to Y_{\infty}$ used the fact that $Y_{n}\subset Y_{n+1}$.
Therefore we embed this sequence by mapping every $Y_{n}$ to some
3-sphere of radius $r$. This mapping is possible by a deep result
of Freedman \cite{Fre:82} showing the embedding of every Casson
handle in the standard open 2-handle given by the interior of a 4-disk.
As shown above, every $Y_{n}$ is part of this Casson handle and by
construction $Y_{n}$ can be embedded into $S^{3}\times(-\epsilon,\epsilon)$
for small $\epsilon$. Then the limit manifold $Y_{\infty}$ must
be mapped to $r=1$, representing the boundary of the Poincare disk
(which is the 'sphere at infinity' for the metric above). Then we
choose that $Y_{1}$ is mapped to the 3-sphere of radius $r=\frac{1}{2}$.
All other $Y_{n}$ have to be mapped in the right order to spheres
of radius between $r=\frac{1}{2}$ and $r=1$. In general, every embedding
is given by a strictly increasing function of the radius $r(n)$ with
respect to the integer $n$. In a particular choice of the embedding,
the number $n$ of the Whitehead double (numbering the 3-manifolds
$Y_{n}$) is related to the radius by 
\begin{equation}
r=1-\frac{1}{n+1}\label{eq:simplest-embedding}
\end{equation}
for the simplest case. In general one has to choose 
\begin{equation}
r=f(n)\quad f(1)=\frac{1}{2},\: f(\infty)=1\label{eq:embedding-condition}
\end{equation}
Finally the embedded $R^{4}$, i.e. $E(R^{4})$, has negative curvature
or (more strongly) admits a hyperbolic 4-metric.

There is also another argument for the negative curvature of the embedded
$R^{4}$ which uses the exotic smoothness structure directly. Namely,
the construction of the small exotic $R^{4}$ refereed to the h-cobordism
directly and leads to the smooth embedding $R^{4}\hookrightarrow S^{4}$
or $R^{4}\hookrightarrow\mathbf{R}^{4}$. A sufficient condition for
this construction is the exoticness of the compact 4-manifolds $X_{+},X_{-}$
which appear in the h-cobordism, i.e. $X_{+},X_{-}$ are homeomorphic
but non-diffeomorphic (see Sec. \ref{sec:Small-exotic}). In the construction
of the example above, it was the pair of compact 4-manifolds with
the topology $E(2)=K3\#\overline{\mathbb{C}P^{2}}$ both ($K3$ for
K3-surface, sometimes also called elliptic surface $E(2)$, see \cite{GomSti:1999})
but non-diffeomorphic as a smooth manifolds. This pair can be simply
constructed by using the Fintushel-Stern knot surgery \cite{FiSt:96},
i.e. one starts with $X_{+}=E(2)$ and modifies $X_{+}$ by using
a knot to obtain $X_{-}$. If the knot has a non-trivial Alexander
polynomial then $X_{+}$ is non-diffeomorphic to $X_{-}$. In our
case, there is a non-trivial solution of the Seiberg-Witten equation
on $X_{-}$ implying the existence of a negatively curved submanifold.
For completeness we will present this argument.

Let us start with the generic case: let $M$ be a compact Riemannian
4-manifold and let $P\to M$ be a Spin$_{C}$ structure for $M$,
i.e. a lift of the principal $SO(4)$ bundle (frame bundle, associated
to the tangent bundle $TM$) to a Spin$_{C}$ principal bundle $P$
with $Spin_{C}=(SU(2)\times SU(2))\times_{\mathbb{Z}_{2}}U(1)$. The
introduction of a Spin$_{C}$ structure do not restrict possible 4-manifolds,
because each 4-manifold admits a Spin$_{C}$ structure. Now one can
introduce the following objects: a Spin$_{C}$ bundle $S_{C}(P)$
as associated vector bundle to $P$ with splitting $S_{C}(P)=S_{C}^{+}(P)\oplus S_{C}^{-}(P)$
(induced by the $\gamma^{5}$ matrix), a canonical line bundle $\det(P)$
with connection $A$, and a Dirac operator $D_{A}:\Gamma(S_{C}^{\pm})\to\Gamma(S_{C}^{\mp})$.
$D_{A}$ depends on the Levi-Civita connection and on the connection
$A$ in $\det(P)$-bundle, and acts on spinors which are sections
$\psi\in\Gamma(S_{C}^{+})$. Then the Seiberg-Witten equations are
formulated as 
\begin{eqnarray*}
F_{A}^{+} & = & q(\psi)=\psi\otimes\psi^{*}-\frac{|\psi|^{2}}{2}id\\
D_{A}\psi & = & 0
\end{eqnarray*}
where $F_{A}=dA$ is the curvature of the connection $A$ and $F_{A}^{+}=\frac{1}{2}(F_{A}-*F_{A})$
is the self-dual part. A trivial (or reducible) solution of these
equations is given by $\psi=0$ (and then $A=d\phi$). To get a non-trivial
Seiberg-Witten invariant, one needs a non-trivial solution $(A,\psi)$
of the equations. For that purpose we consider the second equation
$D_{A}\psi=0$ and obtain for the square of the Dirac operator (Weizenb{{\"o}}ck
formula) 
\[
0=D_{A}D_{A}\psi=\nabla_{A}^{*}\nabla_{A}(\psi)+\frac{R}{4}\psi+\frac{F_{A}^{+}}{2}\psi
\]
where $R$ is the scalar curvature of $M$. Using the first equation
we obtain 
\[
0=\nabla_{A}^{*}\nabla_{A}(\psi)+\frac{R}{4}\psi+\frac{1}{2}\left(\psi\otimes\psi^{*}-\frac{|\psi|^{2}}{2}id\right)\psi
\]
which simplifies to 
\[
0=\nabla_{A}^{*}\nabla_{A}(\psi)+\frac{R}{4}\psi+\frac{|\psi|^{2}}{4}\psi\,.
\]
Now we multiply both sides with $\psi^{*}$ and integrate over $M$
to get 
\begin{equation}
\intop_{M}\left(|\nabla_{A}\psi|^{2}+\frac{R}{4}|\psi|^{2}+\frac{|\psi|^{4}}{4}\right)\sqrt{g}d^{4}x=0\label{eq:SW-relation}
\end{equation}
Except for the scalar curvature term, all other terms are positive
definite. Furthermore, a non-trivial solution of the SW equations
is given by $\psi\neq0$. Therefore the exotic smoothness structure
enforces the manifold $M$ to have a negative scalar curvature $R<0$.
Or, there must exists a submanifold with $R<0$ which dominates the
curvature of $M$. Interestingly, this behavior is unchanged for generic
perturbations (see \cite{Moo:01}).

Now we will use this result by setting $M=X_{-}$. For $X_{+}$ we
can choose the standard smoothness structure allowing only the trivial
solution of the Seiberg-Witten equation ($\psi=0$, $A=d\phi$). By
using the h-cobodism theorem (see section \ref{sec:Small-exotic}),
the h-cobordism between $X_{+}\setminus Y_{+}$ and $X_{-}\setminus Y_{-}$
is trivial and therefore the non-trivial solution of the Seiberg-Witten
equation is localized at $Y_{-}$ and by using $Y_{-}=K\cap R_{-}$
also at $R_{-}$, the small exotic ${R}^{4}$ embedded into $S^{4}$
(or $\mathbf{R}^{4}$). This general argument together with the particular
embedding described above with (\ref{eq:embedding-condition}) gives
the main result of this section:

\emph{The small exotic ${R}^{4}$ embeds into the standard $\mathbf{R}^{4}$
by a smooth map $E:R^{4}\hookrightarrow\mathbf{R}^{4}$. This submanifold
$E(R^{4})\subset\mathbf{R}^{4}$ admits always a negative scalar curvature
$R<0$ and a hyperbolic structure with isometry group $SO(4,1)$.}

\section{Cosmological constant}

So far, we have studied the geometric properties of the embedding
$E:R^{4}\hookrightarrow\mathbf{R}^{4}$ resulting from the topology
of certain infinite chain of 3-manifolds $Y_{1}\to\cdots\to Y_{\infty}$.
As explained above this sequence defines $R^{4}$ as embedded in $\mathbf{R}^{4}$.
Equivalently, the exotic $R^{4}$ is given by an infinite chain of
cobordisms 
\[
End(R^{4})=W(Y_{1},Y_{2})\cup_{Y_{2}}W(Y_{2},Y_{3})\cup\cdots
\]
so that $R^{4}=K\cup_{Y_{1}}End(R^{4})$ (remember $\partial K=Y_{1}$).
Obviously, there is also an embedding $\tilde{E}:End(R^{4})\hookrightarrow\mathbf{R}^{4}$
which will be studied now. The geometry of this embedding is also
given by hyperbolic geometry. This hyperbolic geometry of the cobordism
is best expressed by the metric

\begin{equation}
ds^{2}=dt^{2}-a(t)^{2}h_{ik}dx^{i}dx^{k}\label{eq:FRW-metric}
\end{equation}
also called the Friedmann-Robertson-Walker metric (FRW metric) with
the scaling function $a(t)$ for the (spatial) 3-manifold (denoted
as $\Sigma$ in the following). As explained in the previous section,
this spatial 3-manifold in the case of embedding $E:R^{4}\hookrightarrow\mathbf{R}^{4}$
admits (at least for the pieces) a homogenous metric of constant curvature.

The cosmological constant is defined via Einsteins theory as the eigenvalue
of the Ricci tensor w.r.t. metric, i.e. $Ric=-\Lambda g$. Then one
obtains the equation 
\begin{equation}
\left(\frac{\dot{a}}{a}\right)^{2}=\frac{\Lambda}{3}-\frac{k}{a^{2}}\label{eq:FRW-equation}
\end{equation}
having the solutions $a(t)=a_{0}\sqrt{3|k|/\Lambda}\,\sinh(t\sqrt{\Lambda/3})$
for $k<0$, $a(t)=a_{0}\exp(t\sqrt{\Lambda/3})$ for $k=0$ and $a(t)=a_{0}\sqrt{3|k|/\Lambda}\,\cosh(t\sqrt{\Lambda/3})$
for $k>0$ all with exponential behavior. At first we will consider
this equation for constant topology, i.e. for the spacetime $Y_{n}\times[0,1]$.
But as explained above, the subspace $E(R^{4})\subset\mathbf{R^{4}}$
and every $Y_{n}$ admits a hyperbolic structure. Now taking Mostow-Prasad
rigidity seriously, the scaling function $a(t)$ must be constant,
or $\dot{a}=0$. Therefore we will get 
\begin{equation}
\Lambda=\frac{3k}{a^{2}}\label{eq:CC-to-3D-curvature}
\end{equation}
by using (\ref{eq:FRW-equation}) for the parts of constant topology.
From the topological point of view, $R^{4}$ is homeomorphic to $S^{3}\times[0,1)$
outside of the subset $K\subset R^{4}$ (also called end of $R^{4}$).
From this point of view, we have to choose $k=1/3$ to reflect the
topological $S^{3}$ structure and normalize the curvature so that
$^{3}R=\Lambda$. One can determine the local structure of this space
by considering the isometry groups of the hyperbolic 3- and 4-space.
They are $Isom(\mathbb{H}^{3})=SO(3,1)$ and $Isom(\mathbb{H}^{4})=SO(4,1)$,
respectively. In case of $R^{4}$, the 3-manifolds $Y_{n}$ admit
also hyperbolic structures with isometry group $SO(3,1)$. This group
acts freely on the isometry group of the 4-space $SO(4,1).$ The quotient
space $SO(4,1)/SO(3,1)$ describes the corresponding symmetry reduction
but at the same time, it reflects the local structure of the embedding
$E:R^{4}\hookrightarrow\mathbf{R}^{4}$ . This quotient space is known
as de Sitter space $S^{3}\times[0,1)$ which is compatible with the
embedding, $E$, and its geometry represents physical CC.

Formula (\ref{eq:CC-to-3D-curvature}) can be now written in the form
\begin{equation}
\Lambda=\frac{1}{a^{2}}=^{3}R\label{eq:relation-CC-to-curv}
\end{equation}
so that CC is related to the curvature of the 3D space. This relation
is true for a fixed spatial space but, as explained above, we have
a process $Y_{0}\to\cdots\to Y_{\infty}$ ending at $Y_{\infty}=S^{3}$.
We described the formation of the CC as inflationary process in our
previous work \cite{AsselmeyerKrol2014}. Therefore we define 
\[
\Lambda=curv(Y_{\infty})
\]
by using (\ref{eq:relation-CC-to-curv}), i.e. CC is related to the
curvature of the (wild) 3-sphere surrounding $R^{4}$. Here, there
is an unusual behavior with the following interpretation. We have
a serie of static Einstein universes. Since Eddington \cite{Eddington:1930},
it is known that the static Einstein universe is unstable. This unstability
motivates the series of transitions $Y_{n}\to Y_{n+1}$ in our model
leading to a larger scale of the resulting (spatial) manifold. In
the literature, there is the model of Ellis and Maartens \cite{EllisMaartens:2004}
which have the static Einstein universe in common with our model.
Main difference to our model is the inclusion of topological transitions.
Interestingly, the stability of the static Einstein universe filled
with perfect fluids of different types is currently under investigations,
see for instance \cite{LiWei:2017}.

By using $a_{n}=\sqrt[3]{vol(Y_{n})}$, we are able to define a scaling
parameter for every $Y_{n}$. By Mostow-Prasad rigidity, $a_{n}$
is also constant, $a_{n}=const.$. But the change $Y_{n}\to Y_{n+1}$
increases the volumes of $Y_{n}$, $vol(Y_{n+1})>vol(Y_{n})$, by
adding complements of the Whitehead links. Thus the smoothness structure
of $R^{4}$ determines the spaces (Whitehead link complements) which
have to be added. Therefore we have the strange situation that the
spatial space changes by the addition of new (topologically non-trivial)
spaces. To illustrate the amount of the change, we have to consider
the embedding of $End(R^{4})$ directly. It is given by the embedding
of the Casson handle $CH$ as represented by the corresponding infinite
tree $T_{CH}$. As explained above, this tree must be embedded into
the hyperbolic space. For the tree, it is enough to use a 2D model,
i.e. the hyperbolic space $\mathbb{H}^{2}$. There are many isometric
models of $\mathbb{H}^{2}$(see the appendix \ref{sec:Appendix-models-hyp-geom}
for two models). Above we used the Poincare disk model but now we
will use the half-plane model with the hyperbolic metric 
\begin{equation}
ds^{2}=\frac{dx^{2}+dy^{2}}{y^{2}}\label{eq:hyp-half-plane}
\end{equation}
to simplify the calculations. The infinite tree must be embedded along
$y-$axis and we set $dx=0$. The tree $T_{CH}$, as the representative
for the Casson handle, can be seen as metric space instead of a simplicial
tree. In case of a simplicial tree, one is only interested in the
structure given by the number of levels and branches. The tree $T_{CH}$
as a metric space (so-called $\mathbb{R}-$tree) has the property
that any two points are joined by a unique arc isometric to an interval
in $\mathbb{R}$. Then the embedding of $T_{CH}$ is given by the
identification of the coordinate $y$ with the coordinate of the tree
$a_{T}$ representing the distance from the root. This coordinate
is a real number and we can build the new distance function after
the embedding as 
\[
ds_{T}^{2}=\frac{da_{T}^{2}}{a_{T}^{2}}\quad.
\]
But as discussed above, the tree $T_{CH}$ grows with respect to a
time parameter so that we need to introduce an independent time scale
$t$. From the physics point of view, the time scale describes the
partition of the tree into slices. Then a natural choice seems to
be the setting 
\[
ds_{T}^{2}\sim dt^{2}
\]
where the time scale is related to the hyperbolic distance 
\begin{equation}
\frac{da_{T}^{2}}{a_{T}^{2}}=\frac{1}{L^{2}}dt^{2}\label{eq:generalized-Friedman-equation}
\end{equation}
via the scale $L$ of the hyperbolic structure for the 3-manifold
$Y_{n}$ where $n$ denotes the $n$th level of the tree $T_{CH}$.
The branching of the tree into the level $n$ is related to a length,
denoted by $a_{T}(n)$. Then the level $n$ corresponds to the distance
$a_{T}(n)$ from the root where the tree branches into the $n$th
level of $T_{CH}$. In other words, the tree branches at particular
values of $a_{T}$ but the details are not important for the following
discussion. But this equality is only a heuristic argument. A rigorous
mathematical argumentation is based on the embedding of the tree $T_{CH}$
for a Casson handle $CH$. The Casson handle is a branched surface
and it can be described by quadratic differentials as follows. Let
$X$ be a Riemannian surface then a quadratic differential is a section
of $T^{*}X^{1,0}\otimes T^{*}X^{1,0}$ which is locally given by 
\[
q=q(z)dz^{2}=q(z)\, dz\otimes dz
\]
with the holomorphic function $q(z)$. Away from the zeros of $q(z)$
we can choose a canonical conformal coordinate $\xi(z)=\intop^{z}\sqrt{q}$
so that $q=d\xi^{2}$. Then the set $\left\{ Re(\xi)=const.\right\} $
defines a foliation, called the vertical measured foliation. The holomorphic
function $q(z)$ can be locally expressed as a polynomial. The zeros
of the polynomial are the branching points of the surface, i.e. $z^{n}$
branches into $n$ pieces at $z=0$. Using this result, we are able
to generate the tree $T_{CH}$ by a polynomial. Furthermore we identify
the coordinate $t$ with $t=Re(\xi)$ so that $t=const.$ defines
a vertical foliation (into slices of the constant time). By the deep
theorem of Hubbard and Masur \cite{HubbardMasur:1979}, for every
measured foliation on $X$ (of genus $g>1$) there exists a unique
quadratic differential so that its vertical measured foliation is
equivalent to the measured foliation. In our case, the infinite tree
seen as branched surface is the covering space of a Riemannian surface
of infinite genus. Therefore the quadratic differential is unique
by lifting it to the covering.

In our construction of the $End(R^{4})$, the infinite tree is given
by the process $Y_{0}\to\cdots\to Y_{\infty}$ ending at $Y_{\infty}=S^{3}$.
Every change $Y_{n}\to Y_{n+1}$ defines the branching into the tree
(as given by the branching of the link complements, see sec. \ref{sec:Geometric-properties-of}).
To every $Y_{n}$, one has a hyperbolic structure by choosing a homomorphism
$\pi_{1}(Y_{n})\to SO(3,1)=Isom(\mathbb{H}^{3})$ (up to conjugation).
Because of Mostow-Prasad rigidity, the volume is a topological invariant
and we obtain a natural scale $L$ for every $Y_{n}$. This scale
$L$ changes during the transition $Y_{n}\to Y_{n+1}$ from $L|_{Y_{n}}$
to $L|_{Y_{n+1}}$. But the change is given by the cobordism $W(Y_{n},Y_{n+1})$
and the scales varies as a smooth variable at the cobordism. Therefore,
the foliation with $t=const.$ is enough. One has to consider the
time with respect to the corresponding scale. Therefore we are enforced
to make the identification 
\[
Re(\xi)=\frac{t}{L}
\]
in the quadratic differential (defining the tree $T_{CH}$). Then
the length in the embedded tree is given by $d(Re(\xi$)), the measure
of the vertical foliation. Now the growing $ds_{T}^{2}$ of the tree
with respect to the hyperbolic structure is given by the measure $d(Re(\xi))^{2}$
of the vertical foliation,or 
\[
ds_{T}^{2}=\frac{da_{T}^{2}}{a_{T^{2}}}=d\left(\frac{t}{L}\right)^{2}
\]
in agreement with our heuristic, dimensional argument above. This
equation agrees with the Friedman equation for a (flat) deSitter space,
i.e. the current model of our universe with a CC. This equation can
be formally integrated yielding the expression 
\begin{equation}
a_{T}(t,L)=a_{0}\cdot\exp\left(\frac{t}{L}\right)\label{eq:integration-general-Friedmann}
\end{equation}
and we are enforced to determine the ratio $t/L$. Therefore let 
\[
|{}^{3}R|=\frac{1}{L^{2}}
\]
be the absolute value of the scalar curvature of some $Y_{n}$ with
respect to the scale $L$. By a simple integration with respect to
the metric $h$ of the 3-manifold, $Y_{n}$ we obtain 
\[
\frac{1}{L^{2}}=\frac{\intop_{Y_{n}}\,|^{3}R|\sqrt{h}d^{3}x}{\intop_{Y_{n}}\sqrt{h}d^{3}x}
\]
for the constant scalar curvature. Finally we will show that the integral
is proportional to the Chern-Simons invariant $CS(Y_{n})$.

As a motivation, let us consider the cobordism $W(\Sigma_{1},\Sigma_{2})$
between two 3-manifolds $\Sigma_{1}$ and $\Sigma_{2}$. One important
invariant of a cobordism is the signature $\sigma(W)$, i.e. the number
of positive minus the number of negative eigenvalues of the intersection
form. Using the Hirzebruch signature theorem, it is given by the first
Pontryagin class 
\[
\sigma(W(\Sigma_{1},\Sigma_{2}))=\frac{1}{3}\intop_{W(\Sigma_{1},\Sigma_{2})}tr(R\wedge R)
\]
with the curvature 2-form $R$ of the tangent bundle $TW$. By Stokes
theorem, this expression is given by the difference 
\begin{equation}
\sigma\left(W(\Sigma_{1},\Sigma_{2})\right)=\frac{1}{3}CS(\Sigma_{2})-\frac{1}{3}CS(\Sigma_{1})\label{eq:signature-CS-inv}
\end{equation}
of two boundary integrals where

\[
\intop_{\Sigma}tr\left(A\wedge dA+\frac{2}{3}A\wedge A\wedge A\right)=8\pi^{2}CS(\Sigma)
\]
is known as Chern-Simons invariant of a 3-manifold $\Sigma$. Using
ideas of Witten \cite{Wit:89.2,Wit:89.3,Wit:91.2} we will interpret
the connection $A$ as $ISO(2,1)$ connection. Note that $ISO(2,1)$
is the Lorentz group $SO(3,1)$ by Wigner-In{{\"o}}n{{\"u}} contraction
or the isometry group of the hyperbolic geometry. For that purpose
we choose 
\begin{equation}
A_{i}=\frac{1}{\ell}e_{i}^{a}P_{a}+\omega_{i}^{a}J_{a}\label{eq:Cartan-connection}
\end{equation}
with the length $\ell$ and 1-form $A=A_{i}dx^{i}$ with values in
the Lie algebra $ISO(2,1)$ so that the generators $P_{a},J_{a}$
fulfill the commutation relations 
\[
[J_{a},J_{b}]=\epsilon_{abc}J^{c}\qquad[P_{a},P_{b}]=0\qquad[J_{a},P_{b}]=\epsilon_{abc}P^{c}
\]
with pairings $\langle J_{a},P_{b}\rangle=Tr(J_{a}P_{a})=\delta_{ab}$,
$\langle J_{a},J_{b}\rangle=0=\langle P_{a},P_{b}\rangle$. This choice
was discussed in \cite{Wise2010} in the context of Cartan geometry.
The appearance of the length $\ell$ can be understood by considering
the generators $P_{a}$ and $J_{a}$. $P_{a}$ generates translations
in units of a length, with scale $\ell$, whereas $J_{a}$ generates
rotations in units of an angle. Remember, that every transition $Y_{n}\to Y_{n+1}$
is described by a cobordism $W(Y_{n},Y_{n+1})$. The coordinate normal
to $Y_{n}$ and in direction to $Y_{n+1}$ will be denoted by $t$,
called time. Because of the volume growing $vol(Y_{n})<vol(Y_{n+1})$,
the corresponding length $\ell$ (of the generator $P_{a}$) varies
by every transition $Y_{n}\to Y_{n+1}$ . Then it has the meaning
of the time coordinate parametrizing the transitions. Therefore we
identify $t=\ell$, see below for the consequences. 

Then we obtain for the curvature 
\begin{eqnarray*}
F_{ij} & = & \frac{1}{\ell}P_{a}\left(\partial_{i}e_{j}^{a}-\partial_{j}e_{i}^{a}+\epsilon^{abc}(\omega_{ib}e_{jc}+e_{ib}\omega_{jc})\right)+\\
 &  & J_{a}\left(\partial_{i}\omega_{j}^{a}-\partial_{j}\omega_{i}^{a}+\epsilon^{abc}\omega_{ib}\omega_{jc}\right)
\end{eqnarray*}
In what follows we start with the expression $A\wedge F$ and use
the pairing $\langle\,,\,\rangle=Tr(\,)$ (following the MacDowell
Mansouri approach, see \cite{MacDMan:1977}). Then we will get 
\[
tr(A\wedge F)=\frac{1}{\ell}e\wedge(d\omega+\omega\wedge\omega)+\frac{1}{\ell}\omega\wedge(de+\omega\wedge e)
\]
where the second expression is given by $\omega\wedge T$ with torsion
form $T$. The first expression has the structure $e\wedge R$ with
the curvature 2-form $R$, which agrees with the scalar curvature
$^{3}R$ multiplied by the volume form in the first order formalism.
Therefore for vanishing torsion $T=0$, we obtain 
\[
\intop_{\Sigma}tr(A\wedge F)=\frac{1}{\ell}\intop_{\Sigma}\,^{3}R\sqrt{h}d^{3}x
\]
By using a simple scaling $\omega\to\frac{3}{2}\omega$, we get the
new connection 
\[
\frac{2}{3}A_{i}=\frac{2}{3\ell}e_{i}^{a}P_{a}+\omega_{i}^{a}J_{a}
\]
and finally the relation 
\[
tr\left(A\wedge(dA+\frac{2}{3}A\wedge A)\right)=\frac{3}{2\ell}e\wedge(d\omega+\omega\wedge\omega)
\]
or 
\begin{equation}
8\pi^{2}\cdot\ell\cdot CS(\Sigma)=\frac{3}{2}\intop_{\Sigma}\,^{3}R\sqrt{h}d^{3}x\,.\label{eq:CS-to-EH}
\end{equation}
From (\ref{eq:signature-CS-inv}) it follows that 
\begin{eqnarray*}
\sigma\left(W(\Sigma_{1},\Sigma_{k})\right)=\frac{1}{3}CS(\Sigma_{k})-\frac{1}{3}CS(\Sigma_{k-1})+\frac{1}{3}CS(\Sigma_{k-1})-...-\frac{1}{3}CS(\Sigma_{1})=\\
\frac{1}{3}CS(\Sigma_{k})-\frac{1}{3}CS(\Sigma_{1}).
\end{eqnarray*}
Taking $\Sigma_{2}$ as $\Sigma_{2}=Y_{\infty}$ we can define the
cobordism $End(R^{4})=W(Y_{0},Y_{\infty})$ as $W(\Sigma_{0},\Sigma_{2})$
above, where the limiting 3-manifold $Y_{\infty}$ results from the
resolution of the 1-/2-handle pair by gluing the Casson handle to
$K$ as before.

Thus we will concentrate on the properties of $Y_{\infty}=\Sigma_{2}$
in the following. To relate this cobordism with the $CS$ invariant
as above we need only one assumption: $\Sigma_{2}$ must admit a metric
of constant curvature. After the Thurston's geometrization conjecture
was proved, it is not a strong restriction. $\Sigma_{2}$ must be
a prime manifold with no incompressible torus submanifold. Thus the
formula (\ref{eq:signature-CS-inv}) and the expression above show
that the Chern-Simons invariant of $\Sigma_{2}$ is directly related
to the 4-dimensional cobordism $W(\Sigma_{0},\Sigma_{2})$.

In (\ref{eq:Cartan-connection}) we are enforced to introduce the
length $\ell$ for the translation (represented by the generator $P_{a}$).
But as discussed above, this translation is parametrized by the coordinate
$t$. Above we identify $t=\ell$ with the time and using (\ref{eq:CS-to-EH})
we will obtain the expression 
\begin{equation}
t\cdot CS(\Sigma_{2})=\frac{3}{2}\intop_{\Sigma_{2}}\,^{3}R_{ren}\sqrt{h}d^{3}x\label{eq:scalar-curvature-CS-inv}
\end{equation}
where the extra factor $8\pi^{2}$ (equals $4\cdot vol(S^{3})$) is
the normalization of the curvature integral. This normalization of
the curvature changes the absolute value of the curvature into
\begin{equation}
|^{3}R_{ren}|=\frac{1}{8\pi^{2}L^{2}}\label{eq:renormalized-curvature}
\end{equation}
and we choose the scaling factor by the relation to the volume $L=\sqrt[3]{vol(\Sigma_{2})/(8\pi^{2})}$.
Then we will obtain formally 
\begin{equation}
\intop_{\Sigma_{2}}\,|^{3}R_{ren}|\sqrt{h}\, d^{3}x=\intop_{\Sigma_{2}}\frac{1}{8\pi^{2}L^{2}}\sqrt{h}\, d^{3}x=L^{3}\cdot\frac{1}{L^{2}}=L\label{eq:CS-integral-relation}
\end{equation}
by using 
\[
L^{3}=\frac{vol(\Sigma_{2})}{8\pi^{2}}=\frac{1}{8\pi^{2}}\intop_{\Sigma_{2}}\sqrt{h}\, d^{3}x
\]
in agreement with the normalization above. Let us note that Mostow-Prasad
rigidity enforces us to choose a rescaled formula 
\[
vol_{hyp}(\Sigma_{2})\cdot L^{3}=\frac{1}{8\pi^{2}}\intop_{\Sigma_{2}}\sqrt{h}d^{3}x\,,
\]
with the hyperbolic volume (as a topological invariant). The volume
of all other 3-manifolds can be arbitrarily scaled. In case of hyperbolic
3-manifolds, the scalar curvature $^{3}R<0$ is negative but above
we used the absolute value $|^{3}R|$ in the calculation. Therefore
we have to modify (\ref{eq:scalar-curvature-CS-inv}), i.e. we have
to use the absolute value of the curvature $|^{3}R|$ and of the Chern-Simons
invariant $|CS(\Sigma_{2})|$. By (\ref{eq:scalar-curvature-CS-inv})
and (\ref{eq:CS-integral-relation}) using 
\[
\frac{t}{L}=\begin{cases}
\frac{3}{2\cdot CS(\Sigma_{2})} & \Sigma_{2}\mbox{ non-hyperbolic 3-manifold}\\
\frac{3\cdot vol_{hyp}(\Sigma_{2})}{2\cdot|CS(\Sigma_{2})|} & \Sigma_{2}\mbox{ hyperbolic 3-manifold}
\end{cases}
\]
a simple integration (\ref{eq:integration-general-Friedmann}) gives
the following exponential behavior 
\[
a(t)=a_{0}\cdot e^{t/L}=\begin{cases}
a_{0}\cdot\exp\left(\frac{3}{2\cdot CS(\Sigma_{2})}\right) & \Sigma_{2}\mbox{ non-hyperbolic 3-manifold}\\
a_{0}\cdot\exp\left(\frac{3\cdot vol_{hyp}(\Sigma_{2})}{2\cdot|CS(\Sigma_{2})|}\right) & \Sigma_{2}\mbox{ hyperbolic 3-manifold}\,.
\end{cases}
\]
For the following, we will introduce the shortening 
\[
\vartheta=\begin{cases}
\frac{3}{2\cdot CS(\Sigma_{2})} & \Sigma_{2}\mbox{ non-hyperbolic 3-manifold}\\
\frac{3\cdot vol_{hyp}(\Sigma_{2})}{2\cdot|CS(\Sigma_{2})|} & \Sigma_{2}\mbox{ hyperbolic 3-manifold}
\end{cases}
\]
Finally we can state:\\
 \emph{Let $W(\Sigma_{1},\Sigma_{2})$ be a cobordism which is embedded
into a hyperbolic 4-manifold (succeeding this structure). Let $\Sigma_{1},\Sigma_{2}$
admit metrics of constant curvature. Let the curvature of $\Sigma_{1}$
be $curv(\Sigma_{1})$ (up to a sign), then $\Sigma_{2}$ admits the
curvature 
\begin{equation}
curv(\Sigma_{2})=curv(\Sigma_{1})\cdot\exp(-2\vartheta)\label{eq:cc}
\end{equation}
which we call the cosmological constant for the cobordism. The ratio
of the two curvatures 
\[
\frac{curv(\Sigma_{2})}{curv(\Sigma_{1})}=\exp(-2\vartheta)
\]
is a topological invariant of $\Sigma_{2}$.}

\section{A realistic model for the cosmological constant\label{real-cc}}

Main problem in the calculation of the cosmological constant value
above was the chain of topology changes. In the formula (\ref{eq:cc})
above, we were able to restrict the analysis to the final 3-manifold
$\Sigma_{2}$. Let us consider again the embedding $R^{4}\hookrightarrow\mathbf{R}^{4}$.
We have a chain of 3-manifolds $Y_{0}\to\cdots\to Y_{\infty}$ which
end with $Y_{\infty}$, the wild $S^{3}$. Thus the value of the cosmological
constant induced by the embedding $R^{4}\hookrightarrow\mathbf{R}^{4}$
is given by 
\[
\frac{1}{8\pi^{2}L^{2}}\exp\left(-\frac{3}{CS(Y_{\infty})}\right)
\]
using the normalization (\ref{eq:renormalized-curvature}) and we
need to calculate the Chern-Simons invariant $CS(Y_{\infty})$. But
the Chern-Simons invariant $CS(Y_{\infty})$ vanishes for any 3-sphere
(wild or not). Finally we have:

\emph{The embedding $R^{4}\hookrightarrow\mathbf{R}^{4}$ induces
a vanishing cosmological constant.}

This result is interesting by itself and important for many different
reasons. The example are \cite{AsselmeyerKrol2010,AsselmeyerKrol2011a,AsselmeyerKrol2011d,AsselmeyerMaluga2016}
where we used directly the embedding $R^{4}\hookrightarrow\mathbf{R}^{4}$
and obtained a relation to quantum field theory where the cosmological
constant is given by the vacuum expectation value. This is the embedding
and the hyperbolic geometry which made the short-length modes damped
and the sum over all modes vanishes.

However the vanishing of CC certainly does not explain its small non-zero
value. That is why in the search for the topological origins for the
realistic CC value we look for other, presumably more global, embeddings.
Fortunately, in the case of our small exotic ${R}^{4}$ there is a
natural embedding as given by the construction based on h-cobordism
theorem. Namely, one starts with two non-diffeomorphic but homeomorphic,
compact, simply-connected, closed 4-manifolds $X_{\pm}$ and their
5-dimensional h-cobordism $W$. Now there are submanifolds $Y_{\pm}\subset X_{\pm}$
(called also Akbulut corks) reflecting the non-triviality of the h-cobordism
i.e. the h-cobordism between $X_{+}\setminus Y_{+}$ and $X_{-}\setminus Y_{-}$
is trivial or $X_{+}\setminus Y_{+}$ is diffeomorphic to $X_{-}\setminus Y_{-}$.
Then an open neighborhood of $Y_{\pm}$ inside of the remaining h-cobordism
is the small exotic ${R}^{4}$. It contains the submanifold $K$ having
$Y_{\pm}$ as submanifolds.

In the simplest and well-described example \cite{BizGom:96}, the
small exotic $R^{4}$ was constructed by using the compact 4-manifold
$E(2)\#\overline{\mathbb{C}P^{2}}$ (with the K3 surface $E(2)$)
where the standard $\mathbb{R}^{4}$ is a part of this compact 4-manifold
and serves as the embedding space for $R^{4}$. Therefore we expect
that the real embedding should be finally given by $R^{4}\hookrightarrow E(2)\#\overline{\mathbb{C}P^{2}}$.
Such embedding determines also modified topology changes compared
to the $R^{4}\to\mathbb{R}^{4}$ case. The Akbulut cork of $E(2)\#\overline{\mathbb{C}P^{2}}$
is a contractible 4-manifold with the boundary $\Sigma(2,5,7)$ (a
Brieskorn sphere). Then the small exotic $R^{4}$ can be understood
as lying between the Akbulut cork and the topological part of the
4-manifold $E(2)\#\overline{\mathbb{C}P^{2}}$. Here we will comment
on the similarity of this construction to those from our previous
work \cite{AsselmeyerKrol2014}. In contrast to our previous work
the current model is generic and do not use any special choices (like
a special Casson handle or the embedding of the Akbulut cork). Instead,
it is natural model where $R^{4}$ is the part of a larger spacetime
given by $E(2)\#\overline{\mathbb{C}P^{2}}$ from which $R^{4}$ inherits
its unique data.

As we remember, every subset $K'$, $K'\subset K\subset R^{4}$, is
surrounded by a 3-sphere. Now we take it as Planck-size 3-sphere $S^{3}$
inside of the compact subset $K\subset R^{4}$. This is the initial
point where our cosmos starts to evolve. By the construction of $R^{4}$,
as mentioned above, there exists the homology 3-sphere $\Sigma(2,5,7)$
inside of $K$ which is the boundary of the Akbulut cork for $E(2)\#\overline{\mathbb{C}P^{2}}$.
(see chapter 9, \cite{GomSti:1999}). If $S^{3}$ is the starting
point of the cosmos as above, then $S^{3}\subset\Sigma(2,5,7)$. But
then we will obtain the first topological transition 
\[
S^{3}\to\Sigma(2,5,7)
\]
inside $R^{4}$. The construction of $R^{4}$ was based on the topological
structure of $E(2)$ (the K3 surface). $E(2)$ splits topologically
into a 4-manifold $|E_{8}\oplus E_{8}|$ with intersection form $E_{8}\oplus E_{8}$
(see \cite{GomSti:1999}) and the sum of three copies of $S^{2}\times S^{2}$.
The 4-manifold $|E_{8}\oplus E_{8}|$ has a boundary which is the
sum of two Poincare spheres $P\#P$. Here we used the fact that a
smooth 4-manifold with intersection form $E_{8}$ must have a boundary
(which is the Poincare sphere $P$), otherwise it would contradict
the Donaldson's theorem. Then any closed version of $|E_{8}\oplus E_{8}|$
does not exist and this fact is the reason for the existence of $R^{4}$.
To express it differently, the $R^{4}$ lies between this 3-manifold
$\Sigma(2,5,7)$ and the sum of two Poincare spheres $P\#P$. Therefore
we have two topological transitions resulting from the embedding into
$E(2)\#\overline{\mathbb{C}P^{2}}$ 
\[
S^{3}\stackrel{cork}{\longrightarrow}\Sigma(2,5,7)\stackrel{gluing}{\longrightarrow}P\#P\,.
\]
Each of these two transitions is connected to a different embedding
and therefore we will obtain two contributions (in contrast to the
chain of 3-manifolds for one embedding leading to the single factor
as before). Finally we obtain the two contributions as arranged in
one expression 
\[
a=a_{0}\cdot\exp\left(\frac{3}{2\cdot CS(\partial A_{cork})}+\frac{3}{2\cdot CS(P\#P)}\right)\,.
\]
As mentioned above, the 3-sphere is assumed to be of Planck-size 
\[
a_{0}=L_{P}=\sqrt{\frac{hG}{c^{3}}}
\]
and one obtains for the first transition 
\[
a_{1}=L_{P}\cdot\exp\left(\frac{3}{2\cdot CS(\Sigma(2,5,7)}\right)\,.
\]
To determine the Chern-Simons invariants of the Brieskorn spheres
we can use the method of Fintushel and Stern \cite{FinSte:90,KirKla:90,FreGom:91}.
The calculation can be found in \cite{AsselmeyerKrol2014}. The value
of the Chern-Simons invariant $CS(\Sigma(2,5,7))$ is given by $\frac{9}{280}$
and we obtain 
\[
a_{1}=L_{P}\cdot\exp\left(\frac{140}{3}\right)\approx7.5\cdot10^{-15}m
\]
which is interpreted as the size of the 'cosmos' represented here
by the 3-manifold $\Sigma(2,5,7)$ at the end of the first inflationary
phase \cite{AsselmeyerKrol2014}. This size can be related to an
energy scale by using it as Compton length and one obtains 165 MeV,
comparable to the energy scale of the QCD lying between 217 MeV and
350 MeV (see \cite{EnergyScaleQCD:2004,EnergyScaleQCD:2016}). Thus
starting with the 3-sphere $S^{3}$ of the Planck length size $L_{P}$
and using formula (\ref{eq:renormalized-curvature}) for the renormalized
curvature $1/(8\pi^{2}L_{P}^{2})$ (see the discussion for formula
(\ref{eq:scalar-curvature-CS-inv}) above), the corresponding expression
for the CC reads 
\[
\Lambda=\frac{1}{8\pi^{2}L_{P}^{2}}\cdot\exp\left(-\frac{3}{CS(\Sigma(2,5,7))}-\frac{3}{CS(P\#P)}\right)
\]
which, after introducing the exact values of the Chern-Simons invariants,
$CS(\Sigma(2,5,7)=\frac{9}{280}$ and $CS(P\#P)=\frac{1}{60}$, gives
the value 
\[
\Lambda\cdot L_{P}^{2}=\frac{1}{8\pi^{2}}\exp\left(-\frac{280}{3}-180\right)\approx2.5\cdot10^{-121}
\]
in Planck units.

\emph{Finally we showed that the ratio between cosmological constant
$\Lambda$ and the (normalized) curvature $\Lambda_{BigBang}=\frac{1}{8\pi^{2}L_{P}^{2}}$
of the small $S^{3}$ at the Big Bang is a topological invariant 
\[
\frac{\Lambda}{\Lambda_{BigBang}}=\exp\left(-\frac{3}{CS(\Sigma(2,5,7))}-\frac{3}{CS(P\#P)}\right)
\]
}In cosmology one usually relates the cosmological constant to the
Hubble constant $H_{0}$ (expressing the critical density) leading
to the length scale 
\[
L_{c}^{2}=\frac{c^{2}}{3H_{0}^{2}}\,.
\]
The corresponding variable is denoted by $\Omega_{\Lambda}$ and for
the expression above it gives the topological invariant value 
\begin{equation}
\Omega_{\Lambda}=\frac{c^{5}}{24\pi^{2}hGH_{0}^{2}}\cdot\exp\left(-\frac{3}{CS(\Sigma(2,5,7))}-\frac{3}{CS(P\#P)}\right)\label{eq:dark-energy}
\end{equation}
in units of the critical density. Up to now everything went classical.
The realistic model of CC would certainly require the inclusion of
some kind of quantum corrections to the above calculations. Especially,
the choice of the 3-sphere at the beginning of the Planck size requires
quantum approach. One problem is certainly the lack of any final theory
of quantum gravity. However in the approach to the structure of spacetime
via exotic 4-smoothness (see e.g. \cite{AsselmeyerMaluga2016}) there
exist certain techniques (mainly based on the topology of the handle
decompositions of exotic manifolds), which allow for grasping the
corrections. These corrections are also topological invariants. Namely,
for the first transition, i.e. $S^{3}\to\Sigma(2,5,7)$, we have analyzed
in \cite{AsselmeyerMaluga2016} the corresponding gravitational action.
It appears to be the linear combination of the Pontryagin and Euler
classes of the Akbulut cork of the 5-dimensional non-trivial cobordism
\cite[p. 263]{AsselmeyerMaluga2016}. Because of the contractibility
of the cork, the Pontryagin part has to vanish but the Euler class
gives a nonzero contribution 
\[
\exp\left(-\frac{\chi(A_{cork})}{4}\right)=\exp(-0.25)
\]
with the Euler characteristics $\chi(A_{cork})=1$ of the Akbulut
cork. Introducing this correction into Eq. \ref{eq:dark-energy} the
final formula for $\Omega_{\Lambda}$ reads 
\[
\Omega_{\Lambda}=\frac{c^{5}}{24\pi^{2}hGH_{0}^{2}}\cdot\exp\left(-\frac{3}{CS(\Sigma(2,5,7))}-\frac{3}{CS(P\#P)}-\frac{\chi(A_{cork})}{4}\right)\:.
\]
This additional factor can be also motivated by the short scale behavior
of gravity as shown in \cite{AsselmeyerMaluga2016}. At very small
scales, one obtains a dimensional reduction from 4D to 2D. Then the
4D Einstein-Hilbert action will be reduced to the 2D Einstein-Hilbert
action. But the 2D Einstein-Hilbert action is equal to the Euler characteristics.
In the course of the dimensional reduction, the 4D contractable space
(i.e. the Akbulut cork) will be reduced to a 2D contractable space.
Finally, the 4D Einstein-Hilbert action of the Akbulut cork will be
reduced to the Euler characteristics of the 2D contractable space.
Interestingly, the Euler characteristics for a 2D and a 4D contractable
space agree. This argumentation motivates the appearance of the contribution
$\chi(A_{cork})=1$ in the formula above. 

Current measurements \cite{PlanckCosmoParameters2013,PlanckCosmoParameters2015}
of the Hubble constant in the PLANCK mission combined with other measurements
like the Hubble telescope \cite{HubbleTelescope2016} give the value
\[
\left(H_{0}\right)_{Planck+Hubble}=69,2\,\frac{km}{s\cdot Mpc}
\]
when applied to (\ref{eq:dark-energy}), it gives rise to the following
value of CC 
\[
\Omega_{\Lambda}\approx0.7029
\]
which is in excellent agreement with the measurements. In \cite{AsselmeyerKrol2014}
we discussed some other possibilities of quantum corrections by using
spin foam models or loop quantum gravity. However, the derivation
given above rests on purely topological methods which is favored in
this paper.

\section{Discussion}

We have presented the derivation of the value of the cosmological
constant as a topological invariant based on low dimensional differential
topology. Quantum corrections were also included as topological invariants.
What is the meaning of such topologically supported cosmological constant
derived from the embedding of exotic $R^{4}$ into $E(2)\#\overline{\mathbb{C}P^{2}}$?
One point was mentioned in the Introduction, namely the structure
of the unobserved universe is more rich than the observed local one.
However, everything is still happening entirely in dimension 4, i.e.
$E(2)\#\overline{\mathbb{C}P^{2}}$ is a 4-manifold which local structure
is $\mathbf{R}^{4}$ and exotic smoothness of $R^{n}$ is exclusively
4-dimensional phenomenon. Moreover, we focused on small exotic $R^{4}$
which embeds in the standard $\mathbf{R}^{4}$ as open subset and
it results in the chain of the embeddings $R^{4}\hookrightarrow\mathbf{R}^{4}\hookrightarrow E(2)\#\overline{\mathbb{C}P^{2}}$
explaining global 4-dimensional structure of the universe. One indication
that such structure exists is the topologically supported value of
the dark energy density derived from it. As shown in this paper the
value agrees with Planck mission data. Possibly, some other testable
predictions could be drawn from the existence of such big, nontrivial
structure of the 4-dimensional universe. Among results into this direction
there is the derivation of the speed of the inflation of universe
from exotic smooth structures on $\mathbb{R}^{4}$. The calculations
fit reasonably well with the data of PLANCK \cite{AsselmeyerKrol2014}.

Alternatively, without assuming any big unobserved nontrivial universe
structure, one could still consider the local differentiable structure
of the universe as modeled by smooth exotic $R^{4}$ rather than by
the standard $\mathbf{R}^{4}$ (on which the Lorentz structure is
to be subsequently introduced). This exotic $R^{4}$ can be precisely
defined purely mathematically from the embedding into $E(2)\#\overline{\mathbb{C}P^{2}}$
which follows from the existence of the nontrivial h-cobordism. In
this case one does not assume that universe is described globally
by $E(2)\#\overline{\mathbb{C}P^{2}}$. However, in such a case there
are left open physical questions regarding such choice and the embedding
into the target manifold. On the other hand, this is the embedding
which explains the results obtained and the uniqueness of the Casson
handle structures on which they rest. Thus taking a global structure
of the universe as $E(2)\#\overline{\mathbb{C}P^{2}}$ answers the
questions and gives rise to interesting investigation perspective
where the problem of global gains new meaning. We are leaning towards
and favor the last option in the paper. However, as discussed in \cite{Krol2017}
there are inherent reasons that the exotic $R^{4}$ at cosmological
scales emerged directly from the quantum mechanical structure itself.
In addition exotic $R^{4}$'s can be described by operator algebras
naturally appearing in QM \cite{AsselmeyerMaluga2016}. Hence, if
one agreed that the evolution of the universe began with a quantum
regime where the standard QM description by a Hilbert space of states
applies, the large scale differentiable structure should be an exotic
$R^{4}$ rather than the standard 4-space as shown in \cite{Krol2017}.
Keeping this line of argumentation in mind, we conjecture that this
exotic $R^{4}$ is central for the understanding of the small as well
the large scale structure of the universe. It seems that one could
further attempt to work out this $R^{4}$ directly from QM formalism.
But currently, there is no unique connection. There is ongoing work
on these fascinating topics and near future will yield more definite
resolutions.

\section{Appendix: Hyperbolic 3-/4-Manifolds and Mostow-Prasad rigidity\label{sec:Appendix-Hyperbolic-Mostow}}

In short, Mostow\textendash{}Prasad rigidity theorem states that the
geometry of a complete, finite-volume hyperbolic manifold of dimension
greater than two is uniquely determined by the fundamental group.
The corresponding theorem was proven by Mostow for closed manifolds
and extended by Prasad for finite-volume manifolds with boundary.
In dimension 3, there is also an extension for non-compact manifolds
also called ending lamination theorem. It states that hyperbolic 3-manifolds
with finitely generated fundamental groups are determined by their
topology together with invariants of the ends admitting a kind of
foliation at surfaces in the end. The end of a 3-manifolds has always
the form $S\times[0,1)$ with the compact surfaces $S$. Then a lamination
on the surface $S$ is a closed subset of $S$ that is written as
the disjoint union of geodesics of $S$.

A general formulation of the Mostow-Prasad rigidity theorem is:\\
 Let $M,N$ be compact hyperbolic $n-$manifolds with $n\geq3$. Assume
that $M$ and $N$ have isomorphic fundamental groups. Then the isomorphism
of fundamental groups is induced by a unique isometry.\\
 An important corollary states that \emph{geometric invariants are
topological invariants}. The Mostow-Prasad rigidity theorem has special
formulations for dimension 3 and 4. Both manifolds $M,N$ have to
be homotopy-equivalent and every homotopy-equivalence induces an isometry.
In dimension 3, the homotopy-equivalence of a 3-manifold of non-positive
sectional curvature implies a homeomorphism (a direct consequence
of the geometrization theorem, the exception are only the lens spaces)
and a diffeomorphism (see Moise \cite{Moi:52}). In dimension 4,
compact homotopy-equivalent simply-connected 4-manifolds are homeomorphic
(see Freedman \cite{Fre:82}). This result can be extended to a large
class of compact non-simply connected 4-manifolds (having a good fundamental
group), see \cite{FreQui:90}. Therefore, if a 3- or 4-manifold admits
a hyperbolic structure then this structure is unique up to isometry
and all geometric invariants are topological invariants among them
the volume and the curvature.

Then a hyperbolic 3-manifold $M^{3}$ is given by the quotient space
$\mathbb{H}^{3}/\Gamma$ where $\Gamma\subset Isom(\mathbb{H}^{3})=SO(3,1)$
is a discrete subgroup (Kleinian group) so that $\Gamma\simeq\pi_{1}(M^{3})$.
A hyperbolic structure is a homomorphism $\pi_{1}(M^{3})\to SO(3,1)$
up to conjugacy (inducing the isometry). The analogous result holds
for the hyperbolic 4-manifold which can be written as quotient $\mathbb{H}^{4}/\pi_{1}(M^{4})$.

Let $X^{4}$ be a compact hyperbolic 4-manifold with metric $g_{0}$
and let $M^{4}$ be a compact manifold together with a smooth map
$f:M\to X$. As shown in \cite{BessonCourtoisGallot:1995} or in
the survey \cite{BessonCourtoisGallot:1996} (Main Theorem 1.1),
the volumes of $X,M$ are related 
\[
Vol_{r}(M)\geq deg(f)Vol(X,g_{0})
\]
where $deg(f)$ denotes the degree of $f$. If equality holds, and
if the infimum of the relation is achieved by some metric $g$, then
$(M,g)$ is an isometric Riemannian covering of $(X,g_{0})$ with
covering map $M\to X$ homotopic to $f$. In particular, if $f$ is
the identity map $X\to X$ (having degree $deg(f)=1$) then it implies
that $g_{0}$ is the only Einstein metric on $X$ up to rescalings
and diffeomorphisms.

\section{Appendix: Wild and Tame embeddings\label{sec:Appendix-wild-emb}}

We call a map $f:N\to M$ between two topological manifolds an embedding
if $N$ and $f(N)\subset M$ are homeomorphic to each other. From
the differential-topological point of view, an embedding is a map
$f:N\to M$ with injective differential on each point (an immersion)
and $N$ is diffeomorphic to $f(N)\subset M$. An embedding $i:N\hookrightarrow M$
is \emph{tame}{} if $i(N)$ is represented by a finite polyhedron
homeomorphic to $N$. Otherwise we call the embedding \emph{wild}.
There are famous wild embeddings like Alexanders horned sphere \cite{Alex:24}
or Antoine's necklace. In physics one uses mostly tame embeddings
but as Cannon mentioned in his overview \cite{Can:78}, one needs
wild embeddings to understand the tame one.

\section{Appendix: Models of Hyperbolic geometry\label{sec:Appendix-models-hyp-geom}}

In the following we will describe two main models of hyperbolic geometry
which were used in this paper. For simplicity we will concentrate
on the two-dimensional versions.

The Poincare disk model also called the conformal disk model, is a
model of 2-dimensional hyperbolic geometry in which the points of
the geometry are inside the unit disk, and the straight lines consist
of all segments of circles contained within that disk that are orthogonal
to the boundary of the disk, plus all diameters of the disk. The metric
in this model is given by 
\[
ds^{2}=\frac{dx^{2}+dy^{2}}{(1-(x^{2}+y^{2}))^{2}}
\]
which can be transformed to expression (\ref{eq:metric-hyp-4-ball})
by a radial coordinate transformation. In this model, the hyperbolic
geometry is confined to the unit disk, where the boundary represents
the 'sphere at infinity'.

The Poincare half-plane model is the upper half-plane, denoted by
$\mathbb{H}^{2}=\left\{ (x,y)\:|\: y>0,\: x,y\in\mathbb{R}\right\} $,
together with a metric, the Poincare metric, 
\[
ds^{2}=\frac{dx^{2}+dy^{2}}{y^{2}}
\]
(see (\ref{eq:hyp-half-plane})) that makes it a model of two-dimensional
hyperbolic geometry. Here the line $y=0$ represents the infinity
(so-called ideal points).

Both models are isometric to each other. A point $(x,y)$ in the disk
model maps to the point 
\[
\left(\frac{2x}{x^{2}+(1-y)^{2}},\frac{1-x^{2}-y^{2}}{x^{2}+(1-y)^{2}}\right)
\]
in the half-plane model conversely a point $(x,y)$ in the half-plane
model maps to the point 
\[
\left(\frac{2x}{x^{2}+(1+y)^{2}},\frac{x^{2}+y^{2}-1}{x^{2}+(1+y)^{2}}\right)
\]
in the disk model. This transform is known as Cayley transform.


\section*{References}


\end{document}